\documentclass[twocolumn]{aastex631}

\usepackage{booktabs} 
\usepackage{multirow}
\usepackage{makecell}
\usepackage{tabularx} 
\usepackage[normalem]{ulem} 
\usepackage{color}           

\let\longtable\relax
\let\endlongtable\relax
\let\longtable*\relax
\let\endlongtable*\relax

\shorttitle{Deep Learning-Based Detection and Segmentation of Edge-On and Highly Inclined Galaxies}
\shortauthors{Chrob\'{a}kov\'{a} et al.}

\begin{document}

\title{Deep Learning-Based Detection and Segmentation of Edge-On and Highly Inclined Galaxies}

\author[0000-0002-9895-6638]{\v{Z}. Chrob\'{a}kov\'{a}}
\affiliation{Mullard Space Science Laboratory, University College London, Holmbury St Mary, Dorking, Surrey RH5 6NT, UK}

\author{V. Kre\v{s}\v{n}\'{a}kov\'{a}}
\affiliation{Department of Cybernetics and Artificial Intelligence, Faculty of Electrical Engineering and Informatics, Technical University of Kosice, 042 00 Ko\v{s}ice, Slovakia}

\author{R. Nagy}
\affiliation{Faculty of Mathematics, Physics, and Informatics, Comenius University, Mlynsk\'{a} dolina, 842 48 Bratislava, Slovakia}

\author{J. Gazdov\'{a}}
\affiliation{Department of Cybernetics and Artificial Intelligence, Faculty of Electrical Engineering and Informatics, Technical University of Kosice, 042 00 Ko\v{s}ice, Slovakia}

\author{P. Butka}
\affiliation{Department of Cybernetics and Artificial Intelligence, Faculty of Electrical Engineering and Informatics, Technical University of Kosice, 042 00 Ko\v{s}ice, Slovakia}



\begin{abstract}

Edge-on galaxies have many important applications in galactic astrophysics, but they can be difficult to identify in vast amounts of astronomical data. To facilitate the search for them, we developed a deep learning algorithm designed to identify and extract edge-on galaxies from astronomical images. We utilised a sample of edge-on spiral galaxies from the Galaxy Zoo database, retrieving the corresponding images from the Sloan Digital Sky Survey (SDSS). Our dataset comprised $\sim 16,000$ galaxies, which we used to train the YOLOv5 algorithm for detection purposes. To isolate galaxies from their backgrounds, we trained the SCSS-Net neural network to generate segmentation masks. As a result, our algorithm detected $\sim 12,000$ edge-on galaxies with a high confidence, for which we compiled a catalog including their parameters obtained from the SDSS database.  We described basic properties of our sample, finding that most galaxies have redshifts $0.02<z<0.10$, have low values of $b/a$ and are mostly red, which is expected from edge-on galaxies and is consistent with our training sample, as well as other literature. The cutouts of the detected galaxies can be used for future studies and the algorithm can be applied to data from future surveys as well.

\end{abstract}



\section{Introduction}
Edge-on galaxies are of great interest due to their unique orientation that enables us to study various phenomena. Their high inclination allows for analysis in both radial and vertical direction, which was leveraged since early studies \citep{kormendy_1978,burstein_1979,hamabe_1979,vanderkruit_1981}, usually with only a few well observed galaxies. They can be decomposed into parts, which enables analysis of different galactic components individually \citep[e.g.][and references therein]{tikhonov_2005, mouhcine_2007, comeron_2011,gadotti_2012,salo_2015,mosenkov_2020_decomp,gilhuly_2022}. Edge-on galaxies can be of any type and shape, ranging from bulge-dominated early-type galaxies to poorly understood bulgeless galaxies \citep{kautsch_2006}, which makes them useful for studying various galactic components and peculiar features without biases due to inclination. Some features that are ideally observed in highly inclined galaxies include galactic warps, that were analysed by e.g. \cite{sancisi_1976, reshetnikov_1998, zee_2022}, or galactic fountains, when disk gas is ejected into surrounding medium \citep{bregman_1980, bregman_1997, kluck_2012, putman_2012}. They are also ideal for observations of extraplanar gas \citep[e.g.][and references therein]{reach_2020}, extraplanar dust \citep[e.g.][and references therein]{mosenkov_2022}, polar bulges \citep{corsini_2012,reshetnikov_2015,mosenkov_2024} and other features.

With the increasing volume of data from various surveys, machine learning (ML) is becoming a necessity in galaxy analysis. One of the most common applications of machine and deep learning is for classifying objects into categories. In the galactic context, different ML algorithms are used to classify galaxies based on their morphology into categories on the Hubble sequence (see \citeauthor{huertas_company_2023} \citeyear{huertas_company_2023} for a review) or based on their spectra, which enables categorization of galaxies into different types of active galactic nuclei (AGN), star-forming, composite and other types \citep[e.g.][]{shi_2015, zhang_2019, peruzzi_2021, wu_2024}. Another type of ML application is training algorithms to predict various physical parameters of galaxies, such as photometric redshift (see \citeauthor{soo_2023} \citeyear{soo_2023} for a review) or other physical properties \citep[e.g.][]{bonjean_2019, surana_2020, liew_cain_2021, davidzon_2022, latorre_2024}. ML is also efficient in detecting and extracting particular features of interest such as galactic bars \citep{abraham_2018, cavanagh_2020}, galactic rings \citep{abraham_2025}, tidal features \citep[e.g.][and references therein]{gordon_2024} among others.

The application of ML algorithms specifically for detecting edge-on galaxies has just recently started to get explored. \cite{savchenko_2024} trained an artificial neural network (ANN) on edge-on galaxies extracted from the catalog of genuine edge-on disk galaxies \citep[EGIS,][]{bizyaev_2014}. Then they applied the algorithm on the data from the Panoramic Survey Telescope and Rapid Response System \citep[Pan-STARRS,][]{kaiser_2010}, where they detected $\sim10^5$ galaxies with a detection rate of about $97\%$. \cite{makarov_2022} analyzed the properties of this sample, finding that red sequence galaxies are thicker than the galaxies of the blue cloud and edge-on galaxies are systematically redder than the general population of galaxies seen at arbitrary angles. \cite{usachev_2024} applied this ANN algorithm on images from the Hubble Space Telescope COSMOS field \citep{scoville_2007}, where they detected 950 edge-on galaxies. They studied their scale-lengths and scale-heights, finding that their evolution is luminosity-dependent.

The main motivation of this work is automatic detection and analysis of warped galaxies. In this paper, we present the first step, which is the detection and segmentation of edge-on galaxies, while in a future work we will focus on further training of the algorithm for warped galaxies. This paper is structured as follows. In Section \ref{data}, we describe the selection of images that were used for the algorithm training. In Section \ref{alg}, we explain the processing workflow, algorithms, and its performance in detail. In Section \ref{sample} we present our detected sample and its basic properties. In Section \ref{concl} we summarize our findings.





\section{Data Selection}\label{data}
We are interested in detecting edge-on, spiral galaxies. Galaxies that are not completely edge-on, but are highly inclined are also appropriate, but for simplicity we will refer to our whole sample as edge-on. We selected our original sample from the Galaxy Zoo project \citep{lintott_2008} and we downloaded the primary Galaxy Zoo 2 sample \citep{willett_2013}, which contains classifications of the 243,500 galaxies in the main sample with spectroscopic redshifts. From this sample, following the criteria of \cite{zee_2022}, we selected those galaxies, that at least one half of the volunteers voted to be edge-on and at least 80\% of the volunteers voted to be spiral galaxies. We did not perform any additional visual inspection of the sample. Then we retrieved the fits of these selected galaxies from the The Sloan Digital Sky Survey (SDSS) Data release 7 \citep[DR7]{abazajian_2009} Legacy Survey, which is the final data release of the SDSS Legacy Survey. These FITS images have a size of 2048x1361 pixels, where 1 pixel = 0.396 arcsec. We downloaded them in all $ugriz$ bands, that have magnitude limits $u=22.0$ mag, $g=22.2$ mag, $r=22.2$ mag, $i=21.3$ mag, $z=20.5$ mag. We used all available fits files of all selected galaxies in all filters, so for most galaxies, we have several images. Those are treated as separate objects by the machine learning (ML) algorithm.

\section{Methods}\label{alg}

\begin{figure*}
\centering
\includegraphics[width=\textwidth]{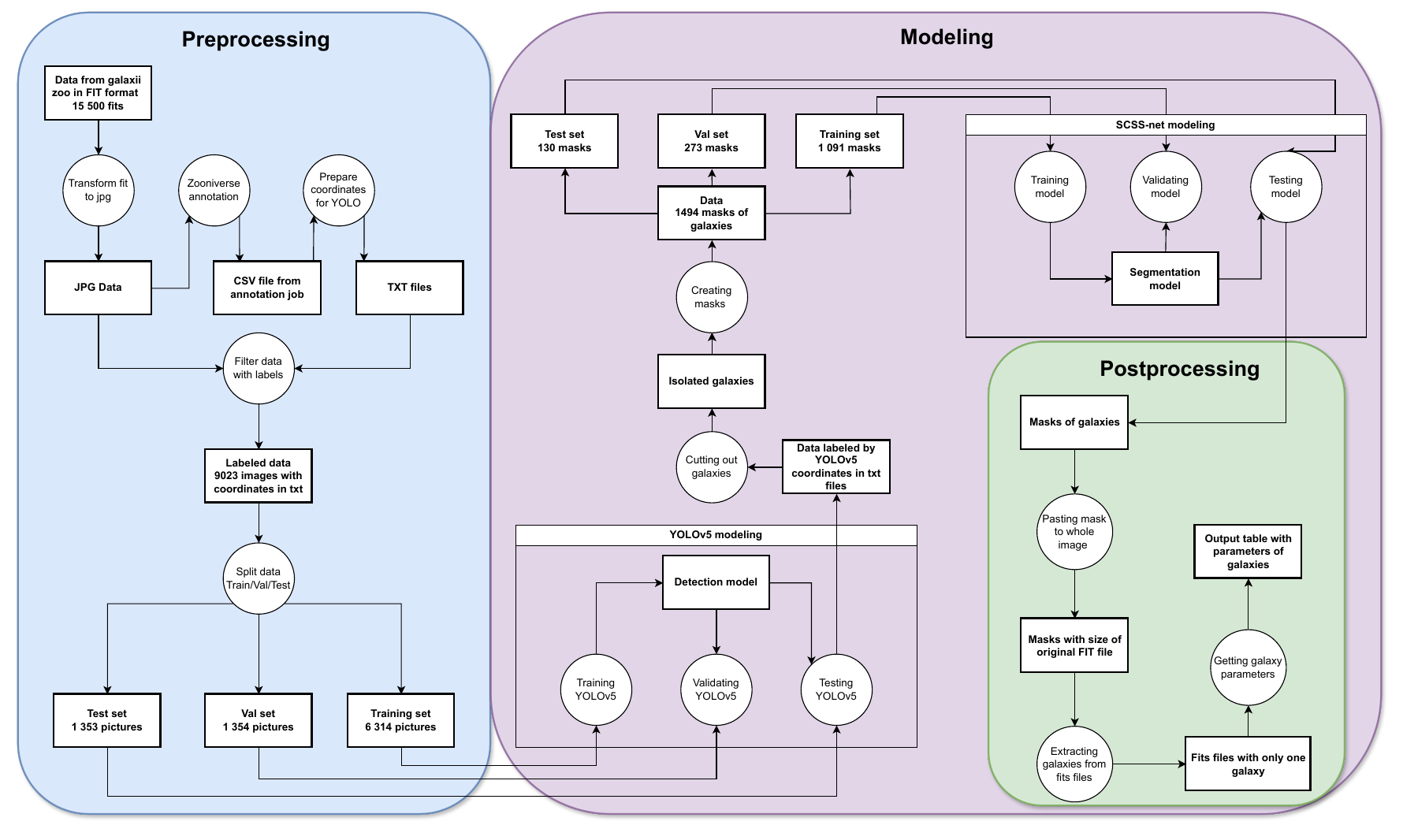}
\caption{Diagram outlining the complete workflow used in the study, illustrating the process of data preparation, modeling and post-processing. Starting with the pre-processing phase, we transform the FITS data into JPEG format and annotate it, followed by splitting it into labeled datasets for machine learning training. The modeling phase involves training the YOLOv5 detection model and the SCSS-net segmentation model to identify and segment galaxies from astronomical images. The post-processing phase focuses on generating segmentation masks and extracting galaxy parameters for further analysis, demonstrating an integrated approach from raw data to usable scientific outputs.}\label{fig:diag}
\end{figure*}

The general overview of our integrated approach is illustrated using the workflow diagram in Fig. \ref{fig:diag}, which depicts efficient processing of galaxy image data to ensure reliable preparation, modeling, and post-processing, developed in this work. This workflow is fundamental to understanding a systematic approach to the analysis of astronomical images and provides a clear path from raw data acquisition to the production of usable scientific data.

The first part of our workflow under Preprocessing block defines the steps that were applied to provide input data for ML algorithms. Within data preparation, we prepared the images in suitable format, annotated them using the Zooniverse platform\footnote{https://www.zooniverse.org/}, and prepared training, validation, and testing sets for the start of the modeling phase. 

The Modeling and Post-processing blocks can be recognized as the main algorithmic part of the processing pipeline and consists of three main procedural steps:
\begin{enumerate}
    \item \textit{Detection} - in this step we acquire detection boxes of edge-on galaxies based on the results from the YOLOv5 detection model, trained on selected data from the SDSS survey. 
    \item \textit{Segmentation} - extraction of cutouts of detected galaxies using retrained U-Net-based architecture.
    \item \textit{Data Extraction} - in this step, segmented masks are used to extract data from fits only for segmented edge-on galaxy, as well as cross-match detected galaxy with SDSS or other catalogs to extract selected parameters. 
\end{enumerate}

We provide more details on all steps of the workflow in the following subsections.

\subsection{Preprocessing and Annotation of Selected Dataset}

Once the images were selected ($15~500$ fits files), they were transformed to JPGs using the grayscale transformation of the FITSFigure wrapper for the transformation of FITS to JPG from \textit{APLpy} package (Astronomical Plotting Library in Python)\footnote{https://aplpy.readthedocs.io/en/stable/}. In order for the images to be useful for annotation, we visualized them using \textit{APLpy's} function \textit{show\_grayscale}, applying a power stretch function with an exponent of $0.5$. The converted images have a size of 1000x727 pixels, with a horizontal and vertical resolution of 100 dpi, and bit depth 24. The images in JPG format were then uploaded to the Zooniverse platform, which became an important component of our data annotation process. The choice of Zooniverse was based on its proven track record of facilitating effective crowd-sourced scientific research, offering tools that simplify the complex tasks of data annotation for users. This platform provided an intuitive interface for volunteers to interact with the data directly, ensuring that the annotations were accurate and verifiable. Galaxy Zoo dataset is a good example of best practice for the use of the platform.

In our case, volunteers were first presented with a brief tutorial explaining the characteristics of edge-on spiral galaxies, as well as how to identify and annotate them. This tutorial included examples of images they would encounter and step-by-step instructions on how to mark the galaxies using the tools provided by the platform.

The project created in Zooniverse was available by public link, but in order to get annotations fast and reliably, we decided to go for annotation session with volunteers from students of Computer Science. We had 3 sessions with approximately 25 students (with 70 different volunteers in total), each 1.5 hours long. Together, the whole annotation process took 4.5 hours. Every session was in the same university classroom with computers and students also had the opportunity to consult any new specific case for annotation, so they were also able to learn during the process on how to proceed correctly with annotations. 

Each image was annotated by one volunteer, so we did not need to take any additional steps to combine more bounding boxes for any image. However, it was possible that in the same image more than one edge-on galaxies occurred. Volunteers were instructed to provide a bounding box for each edge-on galaxy they were able to identify.

\begin{figure*}
\centering
\includegraphics[width=1.0\textwidth]{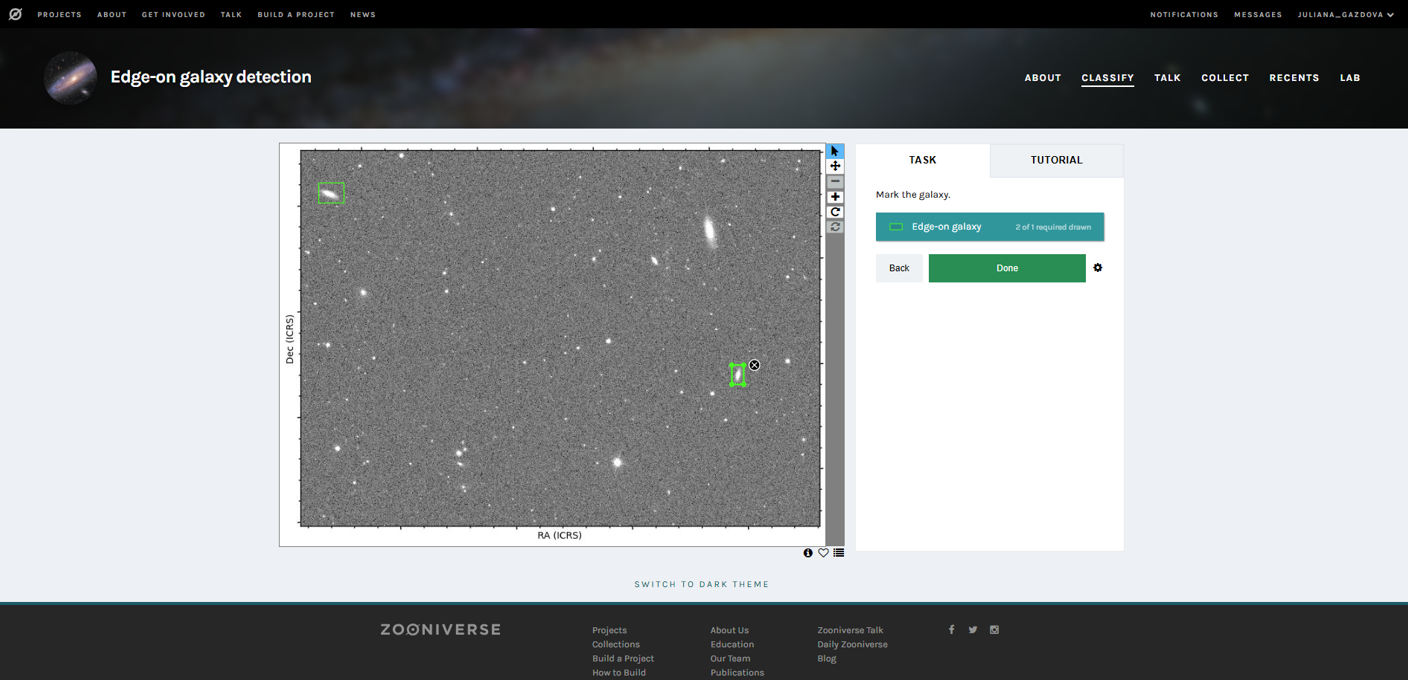}
\caption{An example of an interface with annotation of edge-on galaxies using a green bounding box by a volunteer, already added annotation of second galaxy in current image.}\label{fig:anotation-interface}
\end{figure*}

An example of an annotated image within the Zooniverse project interface is shown in Fig. \ref{fig:anotation-interface}.

The data collected through the annotation step were then compiled and prepared for the next stages of our workflow, focusing on developing a machine learning algorithm to automate the detection of edge-on galaxies. In the end, we had $16~701$ galaxies identified in $9023$ fits files.

Before starting the detection model training, these files were randomly divided into three subsets: training set (70\%), validation set (15\%,) and testing set (15\%).  

\subsection{Detection Task - Setup of YOLOv5 model}

The prepared subsets were used for the training of the detection model, for which we decided to use the YOLO (You Only Look Once) detection network. It is a well-known one-shot (or single-stage) detector, which is used for detection of objects in images, originally developed for real-time monitoring cameras and fast classification of multiple objects in images, with no limit for the number of objects that can be found in one image. We have used the YOLOv5 version of this model \citep{glenn_jocher_2020_4154370}. 

The YOLOv5 detection model follows a structured architecture comprising three main parts: Backbone, Neck, and Head. All details related to the particular implementation of YOLOv5, details on architecture and usage for the own custom datasets can be found in its documentation (see Architecture Summary part in documentation guidelines of YOLOv5)\footnote{https://docs.ultralytics.com/yolov5}. 

The Backbone part extracts features from input images and learns relevant spatial and semantic information using CSPDarknet53, an enhanced version of Darknet53 convolutional network that incorporates Cross Stage Partial (CSP) connections. At the start, YOLOv5 introduces a Focus layer, which reorganizes pixel values to reduce spatial dimensions while preserving important details. In the end of the Backbone part, features of different types and scales are efficiently extracted. The Neck provides a fusion of extracted features, ensuring that information from different scales is effectively combined. YOLOv5 integrates a Path Aggregation Network (PANet), which enhances the flow of information between different feature levels, including the application of the Feature Pyramid Network (FPN) to strengthen multi-scale feature representation. This combination allows the model to detect objects of varying sizes with improved accuracy. The Head part processes these outputs to produce detections (bounding boxes). This architecture ensures efficient and accurate real-time object detection, often used in practice for image processing, making YOLOv5 a robust choice for our galaxy detection step \citep{glenn_jocher_2020_4154370,redmon2016you, vijayakumar2024yolo}.

For our detection training, 6314 images were used to train the model. The validation set with 1354 images helped to evaluate the learning progress of the model during training and optimization of the model. The test set, which includes 1353 images, was used for the final evaluation of the model. It is important to note that if a galaxy appears in different filters, all versions of that galaxy are included in the same subset. This ensures that the same galaxy does not appear in multiple subsets. This condition must be met to maintain the integrity of the data distribution. From a machine learning perspective, the different filters within one subset can be considered as data augmentations, which are very beneficial for including variations in target images for future predictions on unseen data. 

To prepare the data for YOLOv5, we focused on three key columns: \texttt{metadata}, \texttt{subject\_data}, and \texttt{annotations}, extracting essential annotation information. A Python script generated the necessary text files for YOLOv5, containing labels and normalized coordinates $x$ and $y$, where the center of the coordinate system $x,y=(0,0)$ is in the upper left corner of the image. The coordinates were normalized using the following formulas:

\begin{equation}
\text{norm\_x} = \frac{x + \frac{\text{width}}{2}}{\text{picture\_width}}
\end{equation}

\begin{equation}
\text{norm\_y} = \frac{y + \frac{\text{height}}{2}}{\text{picture\_height}}
\end{equation}

Each text file corresponds to an image, with multiple galaxy annotations recorded on separate lines within the same file if present.

We used several key metrics to evaluate the performance of our YOLOv5 model:

\begin{itemize}
    \item \textit{Confusion Matrix}: In general, it is a table or set of numbers which presents the performance of the prediction models by providing number of cases considered as True Positive (TP), False Positive (FP), False Negative (FN) and True Negative (TN) predictions. Due to characteristic and output of detection tasks (TN is considered as background of image without expected detection and is out of scope for such a task), only the first three are applicable and interesting for detection, where: \begin{itemize}
        \item TP is number of cases when object (edge-on galaxy in our case) was annotated for detection and was also correctly detected by the model,
        \item FP is number of cases when object was not annotated, but was detected by the model,
        \item FN is number of cases when object was annotated, but model did not detected it.
    \end{itemize}
    \item \textit{Precision}: The ratio of correctly predicted positive observations (TP) to the total predicted positives (TP and FP), calculated as:
    \begin{equation}
    \text{Precision} = \frac{\text{TP}}{\text{TP} + \text{FP}}
    \end{equation}
    \item \textit{Recall}: The ratio of correctly predicted positive observations (TP) to all observations in the actual class (TP + FN), calculated as:
    \begin{equation}
    \text{Recall} = \frac{\text{TP}}{\text{TP} + \text{FN}}
    \end{equation}

\end{itemize}

Precision and Recall, based on the confusion matrix elements, are metrics which provide comprehensive evaluation of the performance of the model in detecting edge-on galaxies.

\subsection{Segmentation Task Based on U-Net Architecture}

Our main motivation behind this work is the analysis of warps in edge-on galaxies, which we plan to present in future work. Therefore, to analyze our detected galaxies, we need to segment the image and extract the galaxies while removing nearby objects that might contaminate them. There are several methods of image segmentation, such as thresholding, when the object is selected based on having a significantly higher pixel intensity value than the background \citep[e.g.,][]{lotz_2004, zee_2022}, using isophotes to define the extent of the galaxy \citep[e.g.,][]{hao_2006,reshetnikov_2016}, deep learning \citep[e.g.,][]{boucaud_2020, farias_2020} and other methods \citep[for a comprehensive list of methods and their comparisons see][]{xu_2024}. Each method has its advantages and issues, but in all cases determining exactly the threshold where the galaxy ends and the background starts can be challenging.

We decided to extract the galaxies using our own segmentation algorithm. We used the bounding-box coordinates obtained from the YOLOv5 detection to crop galaxies with a close background from the original images. This procedure ensured that the segmented images contained only relevant galaxy data, thus optimizing the segmentation process.

The process of creating high-quality segmentation masks for training the model from cropped galaxy images was time-consuming, mainly due to the manual adjustment of parameters. Each iteration involved careful parameter settings for thresholding and morphological operations such as erosion and dilation. After setting these parameters, we manually checked the resulting masks to ensure that they met our standards for clarity and definition of galaxy structures.

\begin{figure}
\centering
\includegraphics[width=.45\textwidth]{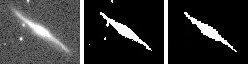}
\caption{Gradual refinement of the segmentation mask for the galaxy. From left to right: Original galaxy image, initial segmentation mask, and refined segmentation mask after iterative parameter adjustment and noise removal.}
\label{fig:tresholding}
\end{figure}

This iterative process of checking, adjusting, and rechecking was necessary to achieve the desired quality of the segmentation masks. The challenges and steps associated with this procedure are shown in Fig. \ref{fig:tresholding}. The goal of each refinement was to remove the remaining irrelevant background noise and improve the edges of the galaxy structures, ensuring that the masks were best suited to train our segmentation models and to get as precise as possible final masks. 

To automate the generation of final segmentation masks, we used SCSS-Net, a convolutional neural network inspired by the U-Net architecture, originally designed to segment solar corona structures \citep{10.1093/mnras/stab2536}. This versatile model was adapted for our specific requirement and re-trained for segmentation of edge-on galaxies.

U-Net is a highly efficient convolutional neural network originally developed for medical image segmentation. Its architecture, designed to work effectively even with a limited amount of data, is characterized by its distinctive U-shape, comprising two main components: an encoder and a decoder \citep{ronneberger2015u}.

\textit{Encoder:} The encoder is responsible for progressively capturing the context of the image, which involves reducing spatial dimensions while increasing depth. This process helps to extract and compress feature information, which is crucial for understanding the complex structures within the image.

\textit{Decoder:} In contrast, the decoder part of the network works by upscaling the feature map to reconstruct the image output. Eventually, it recovers the spatial dimensions and detail of the image, ensuring that the important features extracted by the encoder are used to generate a precise and detailed segmentation. In our case, the input is the original galaxy image with our manually created mask and the output is the mask created by the algorithm. 

\begin{figure*}
\centering
\includegraphics[width=.85\textwidth]{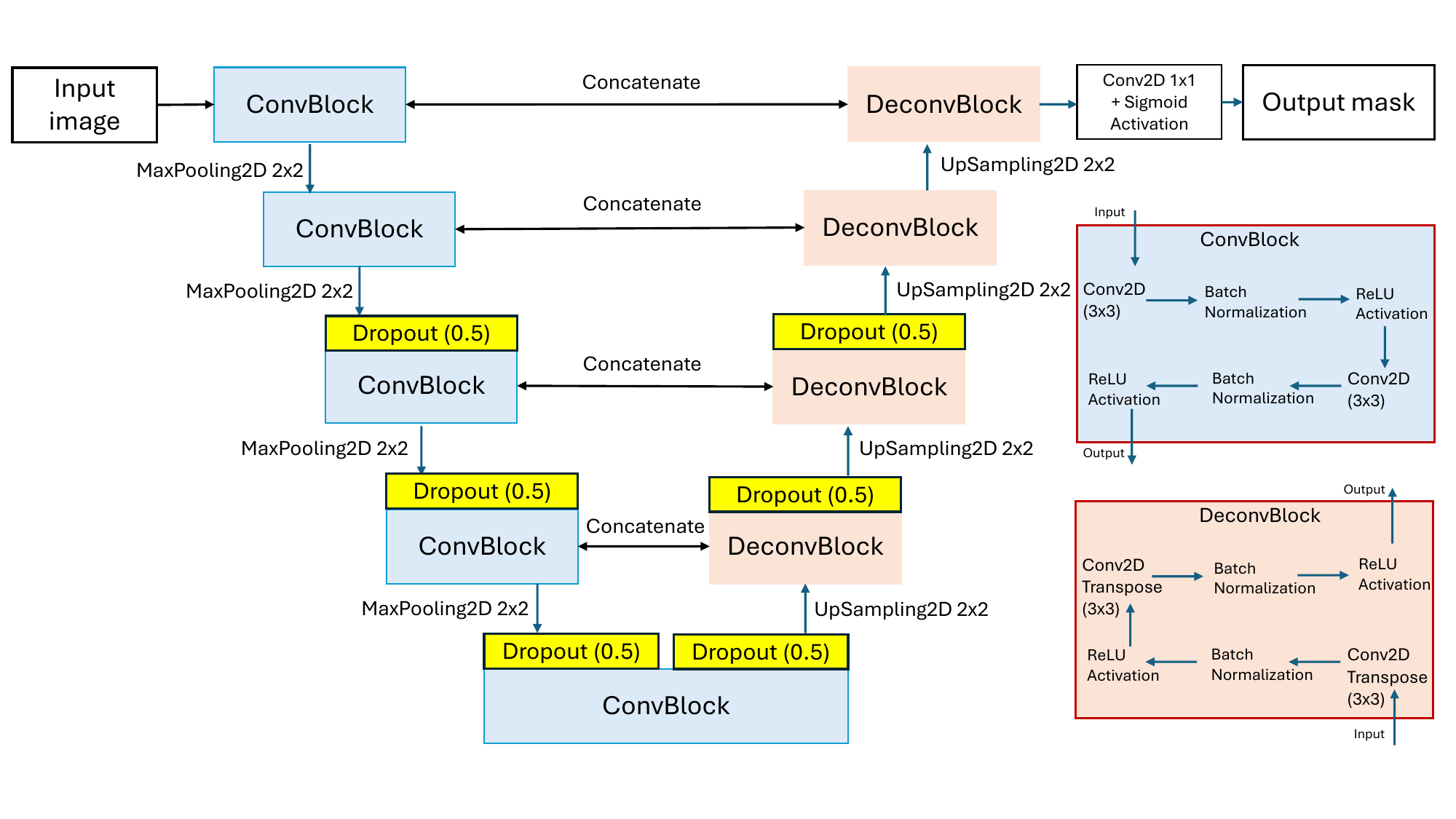}
\caption{The architecture of segmentation model, originally designed for solar corona structures segmentation as SCSS-Net, is inspired by U-Net configuration. For our segmentation of edge-on galaxies, input image (for best model 64x64) of bounding box with edge-on galaxy is at the start of this Encoder-Decoder architecture. The encoder (left part of architecture) is composed of five convolution blocks connected using max pooling layers (with pool size of 2x2 pixels).  Convolutional blocks contain two convolutional layers, batch normalization and ReLU
activation functions (see ConvBlock internal structure depiction on the right side of the figure). The convolutional layers in blocks double their number of filters for each depth level - from 32 at the start to 512 at the bottom of the encoder part. The decoder (right part of the architecture) contains four deconvolutional blocks connected using up-sampling with a pool size of
2×2 pixels. Deconvolution blocks (see DeconvBlock internal structure depicted on the right side of the figure) consist of two transpose convolution layers, batch normalizations, and the ReLU activation functions. Deconvolutional layers progressively halving number of filters from 256 to 32. The deeper layers also contain dropout regularization
layers. Additionaly, concatenation layers are used to concatenate feature maps within the same level of scale (granularity), which 
helps to give the localization information from encoder to decoder. The output layer convolutional layer and sigmoid activation function, providing segmentation output mask for the same size as input image.
}\label{fig:scssnet}
\end{figure*}

A more detailed description of our version of the U-Net model (SCSS-Net) used for edge-on galaxy segmentation is shown in Figure \ref{fig:scssnet}. This architecture not only allows for detailed image segmentation, but also ensures that the network learns from a relatively small amount of data, making it particularly powerful in fields where high-quality data is scarce.

For the learning of the segmentation model, we again divided our data set into training, validation, and test sets, and tuned basic hyperparameters such as image size, batch size, and number of epochs. Due to our previous experience with SCSS-Net, we knew that for this type of task approximately 1000 training images would be sufficient for good segmentation results. Therefore, we prepared enough manually created masks for this purpose. After the preparation of masks, the input dataset with 1494 masks of galaxy images was divided into three subsets: 1091 were used as a training set, 273 masks as validation set, and 130 masks as test set. Due to the small size of our images ($64\times 64$ pixels), the initial training phase was efficient. The whole model was trained from scratch. We tested different image sizes for this purpose (including larger bounding-box inputs, e.g. $128\times 128$ or $256\times 256$), but the best results were achieved using $64\times 64$. We aimed for 1000 epochs with a batch size of 20, but concluded training at epoch 963 as performance gains plateaued.

Accurate evaluation is critical in the development and refinement of image segmentation models. The following metrics are commonly used to assess the quality and accuracy of segmentation results:
\begin{itemize}
    \item \textit{Intersection over Union (IoU):} Also known as the Jaccard Index, IoU is a statistical tool used to quantify the percent overlap between the target mask and the prediction output of the model. It is calculated as the area of overlap between the predicted segmentation and the ground truth divided by the area of union between the predicted and ground truth. This metric is particularly useful for determining how well the predicted boundaries align with the true boundaries.

    \begin{equation}
        \text{IoU} = \frac{\text{Area of Overlap}}{\text{Area of Union}}
    \end{equation}

    \item \textit{Dice Coefficient:} Similar to IoU, the Dice coefficient measures the similarity between two samples. It is defined as twice the area of overlap between the prediction and ground truth, divided by the total number of pixels in both the prediction and ground truth. This makes it a commonly used metric in medical image segmentation due to its sensitivity to small objects.
    
    \begin{equation}
    \text{Dice} = \frac{2 \times \left | \text{Prediction} \cap \text{Ground Truth} \right |}{\left | \text{Prediction} \right | + \left | \text{Ground Truth} \right |}
    \end{equation}
\end{itemize}

\subsection{Evaluation of YOLO-based Detection Results}

The results of the performance of the YOLOv5 model on the test set are summarized in Table \ref{Tab:t1}. The table shows the number of images and its instances of edge-on galaxies, along with the key metrics: Precision and Recall. These results indicate that the model performs well in detecting edge-on galaxies, with a balanced trade-off between precision and recall.

\begin{table}
\centering
\caption{Table with metrics calculated by YOLOv5 on testing dataset. One image may contain more than 1 instance of edge-on galaxy. Precision and Recall are calculated according to True Positive (TP), False Positive (FP) and False Negative (FN) values (see Table \ref{tab:cm}).}
\begin{tabular}{lcccc}
\toprule
Images & Instances & Precision     & Recall    \\ 
\midrule
1353   & 1892      & 0.80 & 0.94 \\ 
\bottomrule
\end{tabular}
\label{Tab:t1}
\end{table}

To provide a comprehensive understanding of the performance of the YOLOv5 model, we also provide particular detection cases, where we can see the sources of errors more precisely. Table \ref{tab:cm} summarizes the counts of true positives (TP), false positives (FP), and false negatives (FN), which are essential to estimate the detection capabilities of the model.

\begin{table}
\centering
\caption{Summary of True Positives (TP), False Positives (FP), and False Negatives (FN) identified by YOLOv5 detection model.}
\begin{tabular}{lccc}
\toprule
Detection case & Count \\ 
\midrule
True Positives (TP) & 1783 \\ 
False Positives (FP) & 442 \\ 
False Negatives (FN) & 113 \\ 
\bottomrule
\end{tabular}
\label{tab:cm}
\end{table}

Due to manual annotations by students, there is some inevitable annotation bias, e.g., not every time students annotated all instances or sometimes a wrong object was annotated. One of the advantages of deep learning techniques is their ability to overcome annotation bias issues and provide correct predictions even for cases with annotation mistakes. This is possible if annotations are generally correct for most cases, and the architecture is able to learn a good generalized prediction function. Hence, the real efficiency of the algorithm is even better and might be additionally estimated by the analysis of wrong cases.

\begin{figure}
\centering
\includegraphics[width=.45\textwidth]{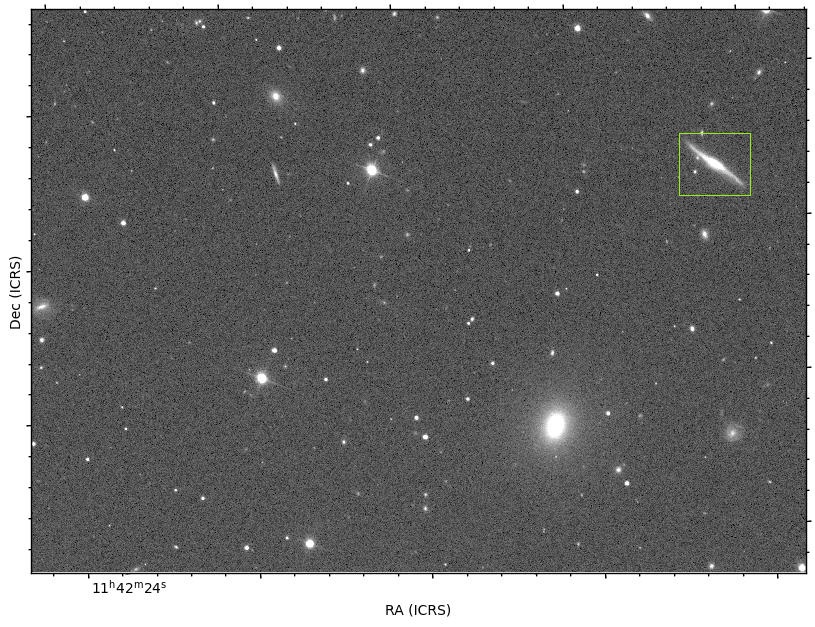}
\caption{An example of an edge-on spiral galaxy annotated with a green bounding box by a volunteer.}\label{fig:anotation}
\end{figure}

\begin{figure}
\centering
\includegraphics[width=.45\textwidth]{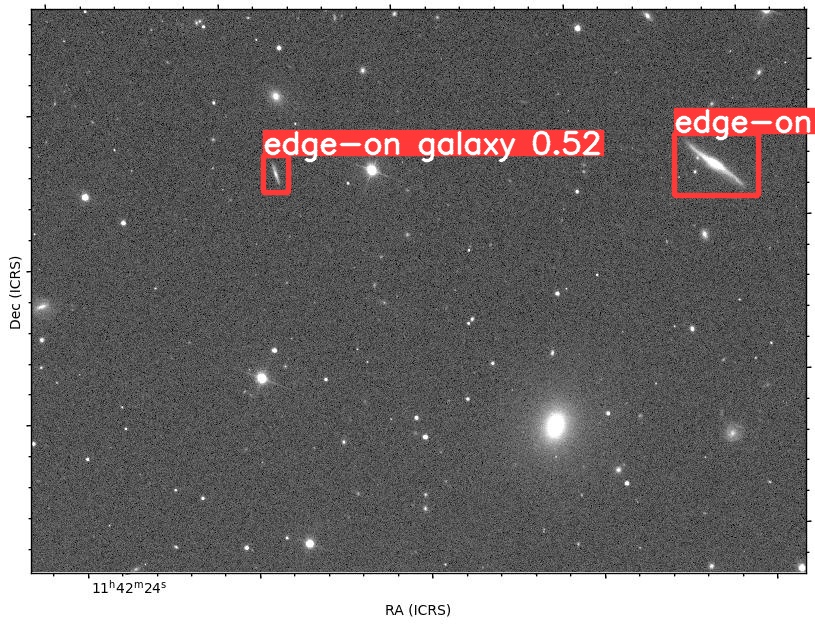}
\caption{The same image as Fig.\ref{fig:anotation}, but showing the galaxies detected by the YOLOv5 detection model. There is one extra galaxy identified, compared to the annotation.}
\label{fig:yolo_annotation}
\end{figure}

According to our knowledge from the Galaxy Zoo project, there should be at least one edge-on galaxy in the image, but that does not rule out the possibility of detecting other galaxies. Initially, our annotators only flagged galaxies that they were confident of, but our algorithm successfully detected several others. This is illustrated in Fig. \ref{fig:anotation}, where we present an example image that initially had one galaxy annotation from Zooniverse. As can be seen in Fig. \ref{fig:yolo_annotation}, after applying a trained YOLOv5 model, an additional galaxy was detected. This demonstrates the capability of the model to identify galaxies that may have been missed in the initial annotations due to manual annotation bias.

\begin{figure*}
    \centering
    \includegraphics[width=\textwidth]{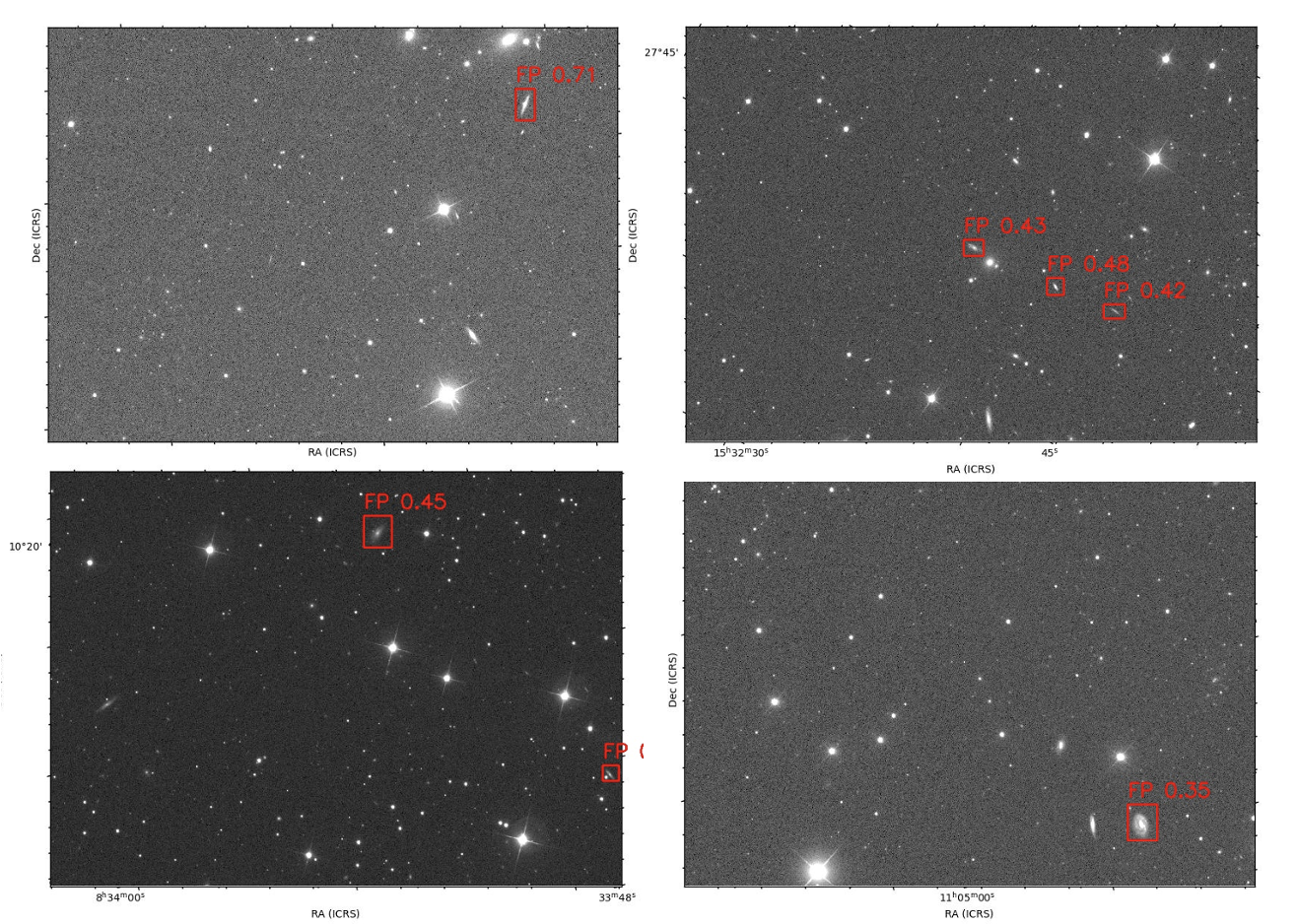}
    \caption{Examples of False Positives Detected by YOLOv5. We can see that the model's ability to identify galaxies that were often missed in the initial annotations and these objects might be considered as True Positives instead of False Positives.}
    \label{fig:fp}
\end{figure*}

\begin{figure*}
    \centering
    \includegraphics[width=\textwidth]{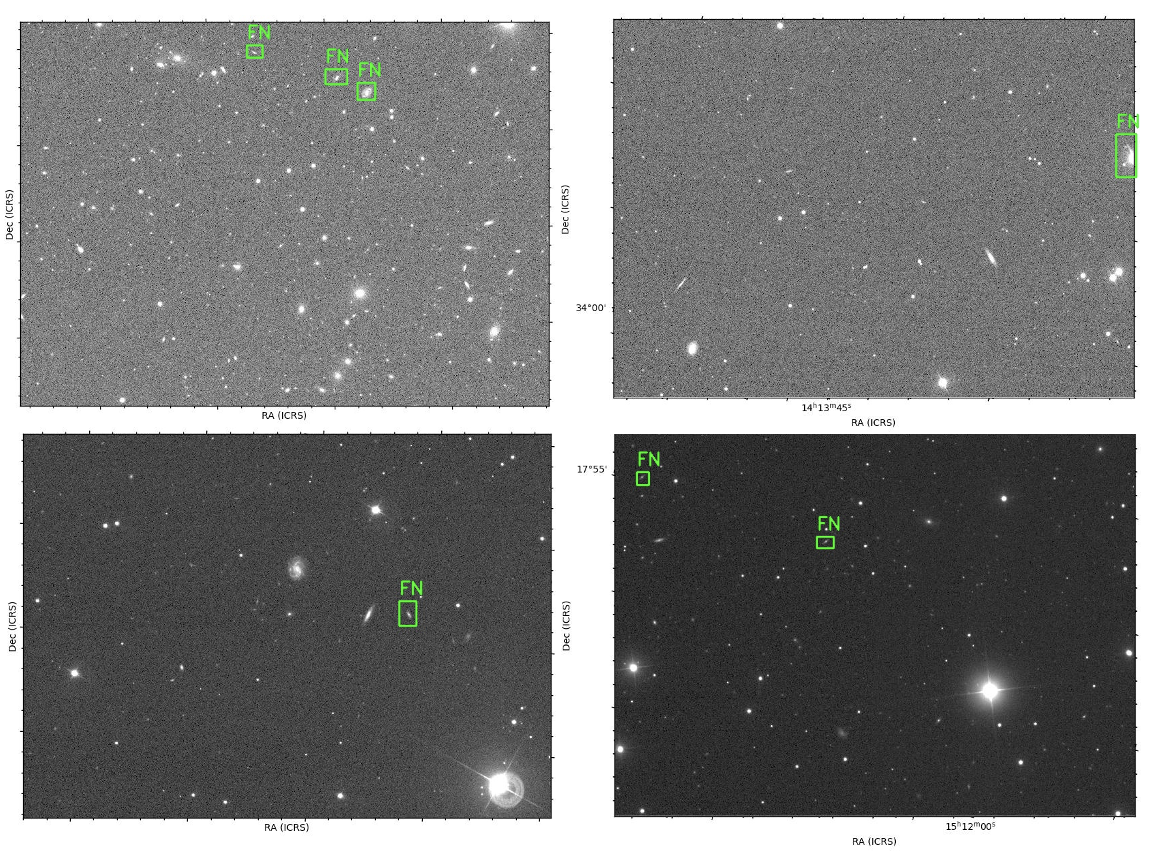}
    \caption{Examples of False Negatives Detected by YOLOv5. These often represent stars, incorrectly annotated by the annotators, confirming their classification as true negatives, which in reality do not contribute to our detection metrics.}
    \label{fig:fn}
\end{figure*}

To verify these findings, we manually examined detections labeled false positives (FP) and false negatives (FN). Figures \ref{fig:fp} and \ref{fig:fn} show cases of false positives and false negatives, providing visual context for these detections.

Through visual inspection, we found that a number of misclassifications are due to inaccuracies in the annotation. In more detail, at least $\sim 20\%$ of the FP cases are considered edge-on galaxies, which were not annotated by mistake, and therefore these cases might be relocated from FP to TP after such analysis. For another $\sim 20\%$ of galaxies we were not able to decide, based on visual inspection, what their inclination is, but some of them might be edge-on as well. Hence, we estimate that the precision is probably at least $\sim 85\%$ in comparison to $\sim 80\%$ measured for the original annotations. Although these numbers are only approximate, as they were obtained by visual inspection, which is prone to bias, they suggest that performance indicators may be underestimated and could potentially be higher with more accurate initial data. This observation highlights one of the key strengths of deep learning models mentioned before: the ability to learn the correct prediction function, even in the presence of partially imperfect annotations.

\subsection{Evaluation of U-Net Segmentation Results}

Testing revealed an average IoU of 0.75 and a Dice coefficient of 0.86, reflecting the efficiency of the model. Bright stars within the images were the main source of errors, a challenge we plan to address in the future. Despite these issues, the segmentation results were predominantly precise and well-defined. In segmentation tasks, we often see lower values of IoU and Dice score, which is caused by the fact that ground truth masks are manually or semi-manually produced (using some setup of selected tresholding and morphological operations), i.e., the ground truth masks are not ideal, but instead have some bias. However, deep learning is able to overcome such bias thanks to its learning capacity with more complex function defined by weights in its architecture and provide better segmentation results. Therefore, IoU and Dice will not reach optimal values, as they are measuring overlap with biased ground truth masks, but will be of relatively high value. In such cases, it is also important to do qualitative checks for final segmented masks and see whether these masks are mostly valid.  

\begin{figure*}
\centering
\includegraphics[width=\textwidth]{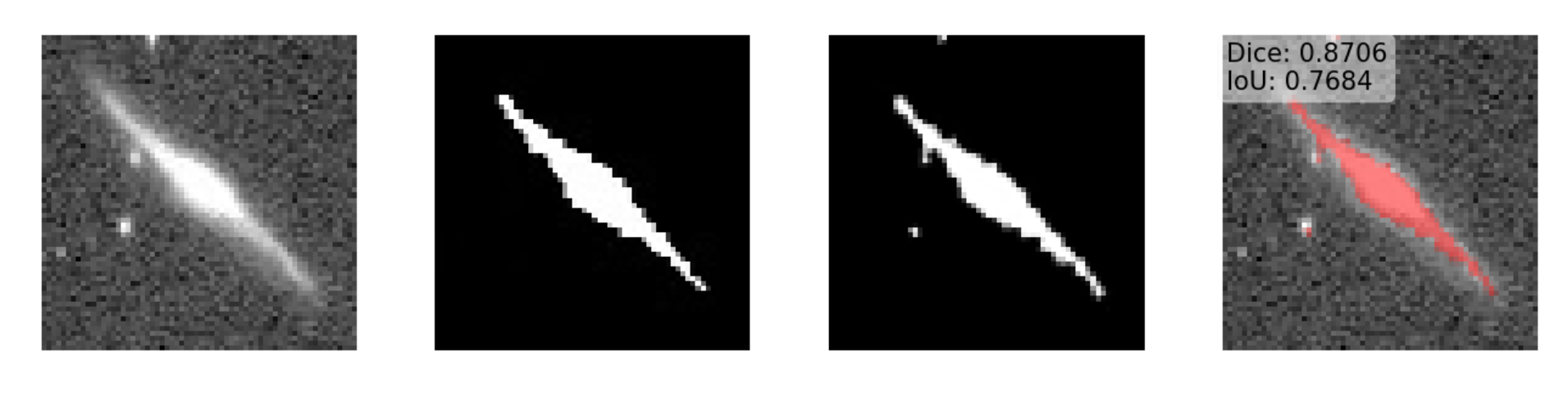}
\caption{Detailed stages of galaxy image segmentation. From left to right: the original galaxy image, the ground truth mask, the segmentation result, and the final overlay highlighting the segmentation accuracy with performance metrics.}
\label{fig:galaxy_segmentation_process}
\end{figure*}

As illustrated in Fig. \ref{fig:galaxy_segmentation_process}, the segmentation process is detailed through four sequential stages. The first panel shows the original galaxy image, followed by the ground truth mask in the second panel, which serves as the training and validation sample for the performance of our model. The third panel displays the result of the segmentation produced by our model, and the final panel overlays this segmentation on the original image to highlight the model’s accuracy and effectiveness. Notably, the Dice coefficient and IoU values, 0.87 and 0.77 respectively, demonstrate the high precision and effectiveness of our segmentation approach.

\begin{figure*}
\centering
\includegraphics[width=0.97\textwidth]{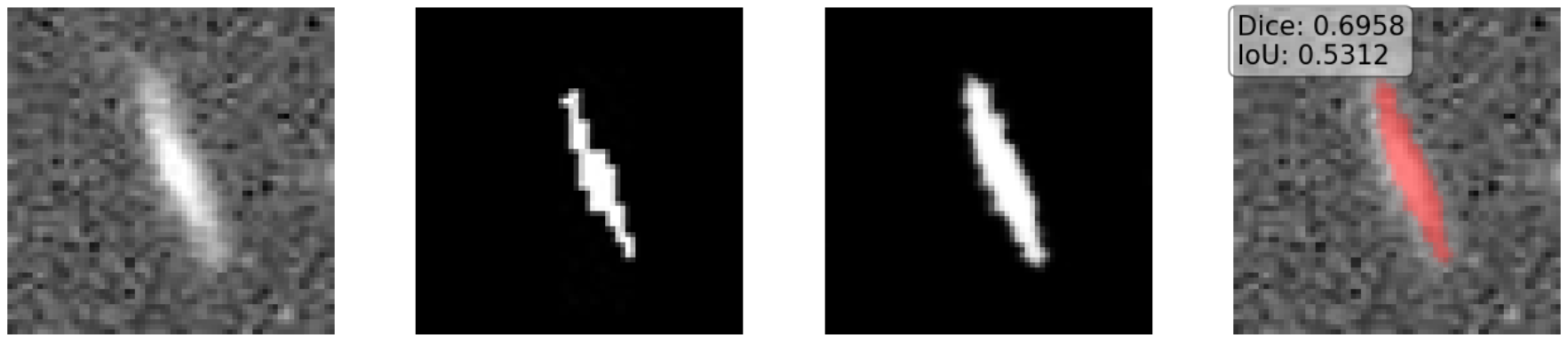}
\caption{Comparison of manual and automated segmentation techniques for a galaxy image. From left to right: the original galaxy image, the manual ground truth mask, the automated segmentation result, and an overlay highlighting the segmentation accuracy. The Dice coefficient is 0.6958 and the IoU is 0.5312, indicating that the automated segmentation outperforms the manual mask in defining galaxy structures.}
\label{fig:improved_segmentation}
\end{figure*}

The model demonstrated reliability in its discrimination of relevant pixels for segmentation, often outperforming the quality of our manual masks, as reflected in the metrics. An example of such segmentation and its validation is shown in Fig. \ref{fig:improved_segmentation}. We found that combination of generally high scoring metrics and visual inspection of segmented edge-on galaxies showed good and stable results for final setup of algorithm.  

\begin{deluxetable*}{cccccccccccc}
    \tablecaption{Selected parameters of the detected sample, queried from the SDSS database, unless stated otherwise. The full version of the table is available at our Github repository \url{https://github.com/MatFyzGalaxy/EdgeOnML}. Columns: (1): Unique SDSS identifier, (2): Galaxy name from NED, (3) and (4): J2000 coordinates, (5)-(9): Petrosian flux in the respective bands, (10): Exponential fit \ensuremath{b/a} in the $r$ band, (11): Redshift, (12): Confidence with which the galaxy was detected, given by the algorithm.}
    \tablewidth{0pt}
    \tabletypesize{\scriptsize}
    \colnumbers
    \tablehead{
        \colhead{$ID$} & \colhead{$Name$} & \colhead{\makecell{$RA$ \\$[deg]$}} & \colhead{\makecell{$DEC$ \\ $[deg]$}} & \colhead{\makecell{$u$ \\ $[mag]$}} & \colhead{\makecell{$g$ \\ $[mag]$}} & \colhead{\makecell{$r$ \\ $[mag]$}} & \colhead{\makecell{$i$ \\ $[mag]$}} & \colhead{\makecell{$z$ \\ $[mag]$}} & \colhead{$(b/a)_r$} & \colhead{$Redshift$} & \colhead{$Confidence$}
    }
    \startdata
        1237663237128519791 & UGC 02699 & 50.5042  & -1.0535  & 18.281 & 16.143 & 15.209 & 14.738 & 14.324 & 0.275 & 0.0266 & 0.861 \\
        1237665369582928005 & UGC 06539 & 173.3493 & 32.6037  & 21.547 & 21.176 & 21.645 & 22.773 & 22.827 & 0.999 & -      & 0.859 \\
        1237648705119060162 & FGC 1534 & 195.1273 & 0.4801   & 17.952 & 16.648 & 15.897 & 15.418 & 15.067 & 0.190 & 0.0682 & 0.840 \\
        1237667730200264899 & UGC 04742 & 135.5952 & 14.5252  & 17.095 & 15.317 & 14.437 & 13.969 & 13.611 & 0.444 & 0.0301 & 0.839 \\
        1237661358611824658 & CGCG 241-036 & 163.6793 & 46.5540  & 16.461 & 14.683 & 13.828 & 13.387 & 13.070 & 0.485 & 0.0222 & 0.837 \\
        1237655126084878404 & VCC 1597 & 188.7607 & 5.4249   & 17.688 & 16.175 & 15.533 & 15.219 & 15.284 & 0.204 & -      & 0.836 \\
        1237667734496084000 & UGC 05535 & 153.9388 & 18.9461  & 16.652 & 14.914 & 14.047 & 13.617 & 13.315 & 0.477 & 0.0216 & 0.835 \\
        1237667322722844715 & UGC 07959 & 191.9249 & 26.9816  & 16.183 & 14.344 & 13.510 & 13.096 & 12.789 & 0.473 & 0.0239 & 0.834 \\
        1237663785276670105 & FGC 0096 & 12.7149  & 0.8513   & 17.906 & 16.511 & 16.104 & 15.839 & 15.817 & 0.123 & 0.0153 & 0.833 \\
        1237657606426263745 & CGCG 208-032 & 129.4910 & 39.8962  & 17.073 & 15.478 & 14.642 & 14.208 & 13.823 & 0.426 & 0.0242 & 0.832 \\
    \enddata
    \label{tab:gal_parameters}
\end{deluxetable*}

\begin{figure}
    \centering
    \includegraphics[width=0.45\textwidth]{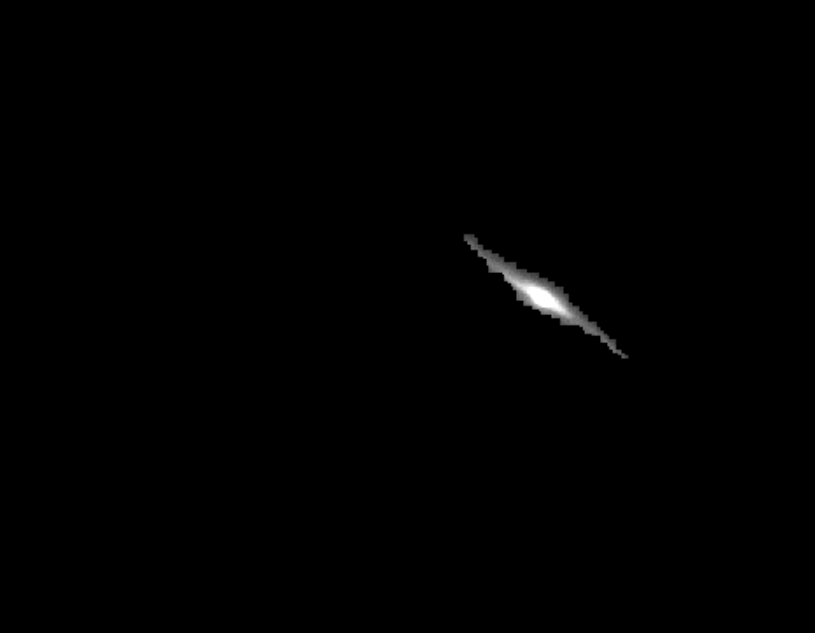}
    \caption{Illustration of image data extraction from FITS files, showing the accuracy and efficiency of the process.}
    \label{fig:fits}
\end{figure}

After advanced segmentation techniques, the next key step was to extract relevant data from the FITS files using the resulting masks. This process is crucial to process our galaxy images for further understanding and analysis. The result of this extraction can be seen in Fig. \ref{fig:fits}. This approach allows us to efficiently search existing galaxy databases and use the compiled data for a comprehensive analysis, which is detailed in the next section.

To ensure complete transparency of our process and facilitate reproducibility, we provide comprehensive documentation, including all codes used in our analyses, available on our GitHub repository\footnote{\url{https://github.com/MatFyzGalaxy/EdgeOnML}}.



\section{Properties of the selected sample}\label{sample}
In this section, we show some basic properties of the sample detected by our algorithm and show that in general they are of the same type of galaxies that we trained the algorithm on. The finished algorithm was run on the full SDSS DR17 sample \citep{abdurrouf_2017}, where we detected $171,946$ objects. However, as we discuss below, many of these are detected with a low confidence as they are faint and blurry, so in further analysis we only consider galaxies detected with $Confidence > 0.7$, which results in about $\sim 12,000$ galaxies.

After segmentation and extraction of the galaxies from their fits files, we queried their coordinates in the SDSS catalog, retrieving their redshifts, magnitudes and $b/a$ ratio using the \textit{SciServer} package. We also retrieved the names of the galaxies from the NED database using the \textit{astroquery} package \citep{ginsburg_2019}. In Table \ref{tab:gal_parameters} we show the parameters of the first ten galaxies from our sample. The full table is available at our Github repository as well.

Then we calculate the absolute magnitudes using the usual formula
\begin{equation}
M=m-25-5\mathrm{log}_{10}(d(1 + z))-K~,
\end{equation}
where $m$ is the Petrosian magnitude, corrected for the Galactic extinction using the map of \cite{schlafly_2011}, $d$ is the comoving distance in Mpc, calculated using Python's \textit{astropy} package assuming flat $\Lambda$CDM cosmological model with total matter density and dark energy density $\Omega_m=0.27, \Omega_\Lambda=0.73$ respectively and Hubble constant $H_0=70$ km s$^{-1}$ Mpc$^{-1}$. $K$ is the $k+e$ correction, where the $k-$correction was calculated using the calculator \footnote{\url{http://kcor.sai.msu.ru/}} based on \cite{chilingarian_2010} and \cite{chilingarian_2012} and the evolution correction was determined using the expression
\begin{equation}
E(z)=Q_0[1+Q_1(z-z_0)](z-z_0)
\end{equation}
from \cite{blanton_2006}, where $Q_0=2.8$, $Q_1=-1.8$ and $z_0=0.1$. This formula was derived for the SDSS sample, with redshifts $0.05<z<0.15$, which is satisfied for our sample.

\begin{figure*}
\centering
\includegraphics[width=\textwidth]{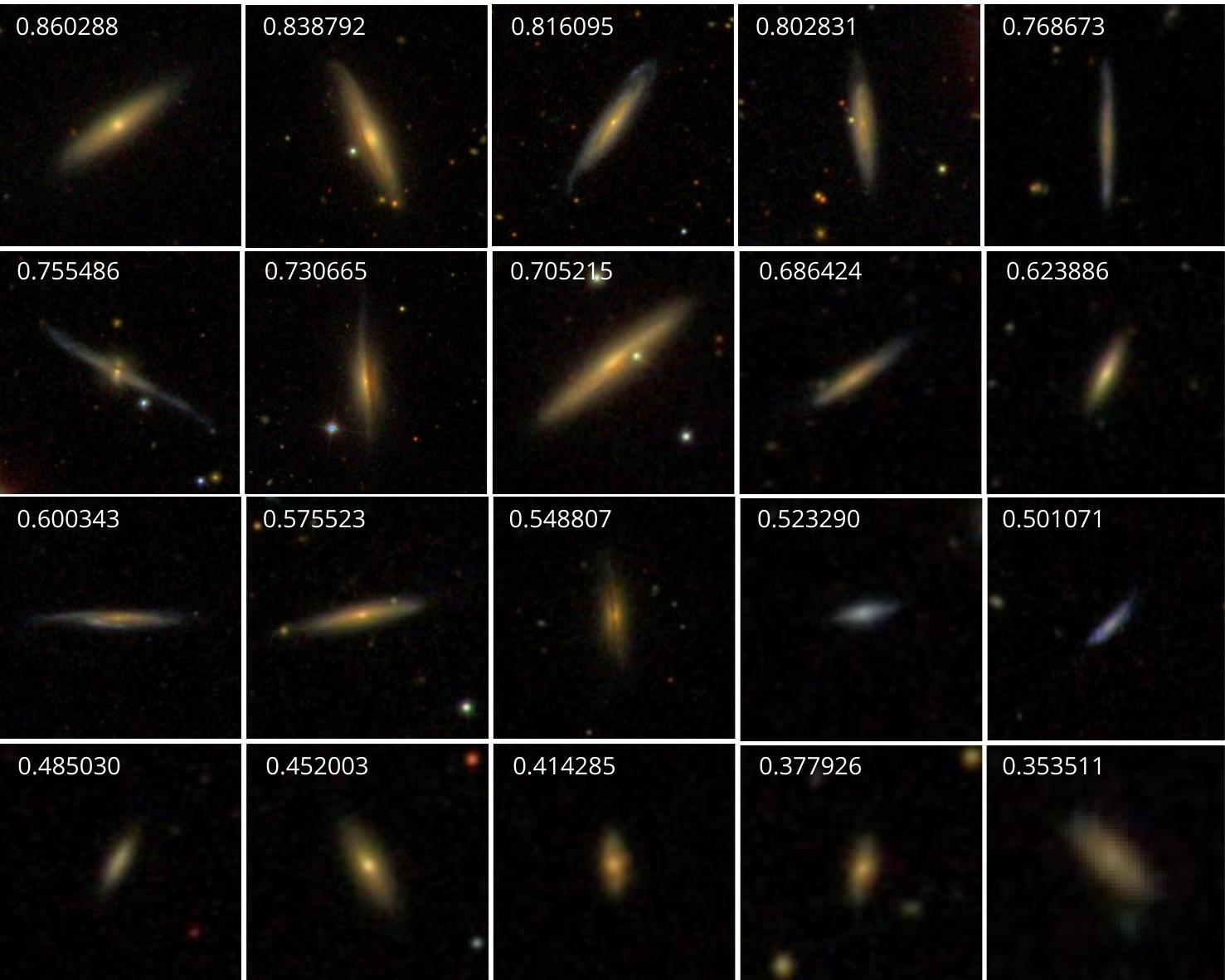}
\caption{Examples of randomly selected detected galaxies, with various confidence values.}\label{f:5}
\end{figure*}

In Fig. \ref{f:5} we show an example of randomly selected galaxies with the respective confidence given by the algorithm. By inspecting the galaxies visually, we can see that the algorithm is rather conservative, and galaxies that are clearly edge-on or highly inclined are detected with $0.7 < Confidence < 0.9$. Galaxies with $Confidence < 0.5$ are mostly small and/or faint, therefore, their nature could be debated. The default confidence threshold for YOLOv5 is set to $0.25$. This value determines the minimum confidence score required for the model to consider a detected object as valid. Adjusting this threshold can help balance between detecting all possible objects (sensitivity) and minimizing false positives (precision). After optimising on our validation dataset, we used an optimal confidence threshold of $0.3$ to achieve a better balance between precision and recall, ensuring more reliable detections in our specific context. Therefore, we provide all galaxies detected with $Confidence > 0.3$, but for further astrophysical analysis, we recommend using galaxies with at least $Confidence > 0.5$ or higher. In the following analysis, we only consider galaxies with $Confidence > 0.7$.

\begin{figure}
    \centering
    \includegraphics[width=0.45\textwidth]{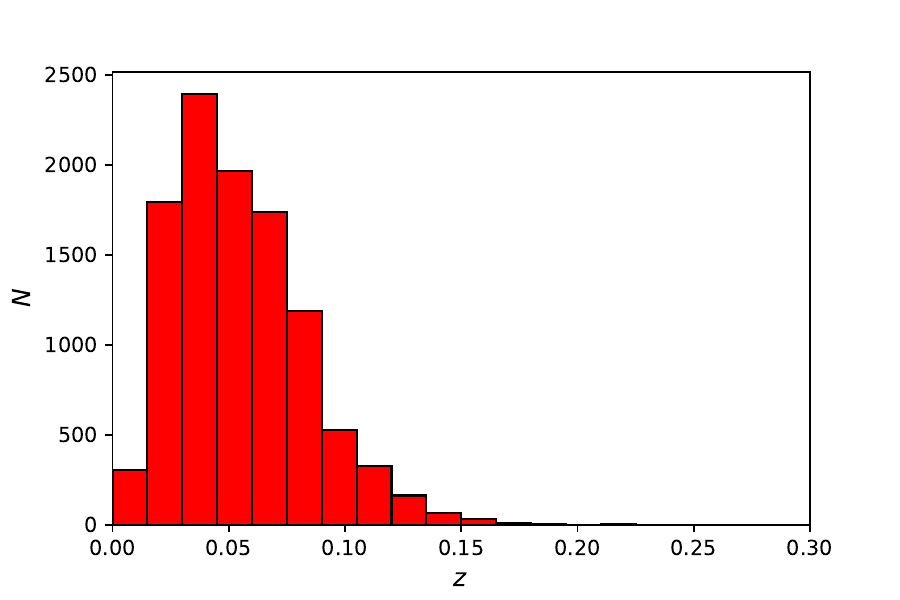}
    \caption{Distribution of redshifts of the detected sample with $Confidence>0.7$ are considered.}
    \label{f:2}
\end{figure}

\begin{figure*}[!b]
    \centering
    \includegraphics[width=0.36\textwidth]{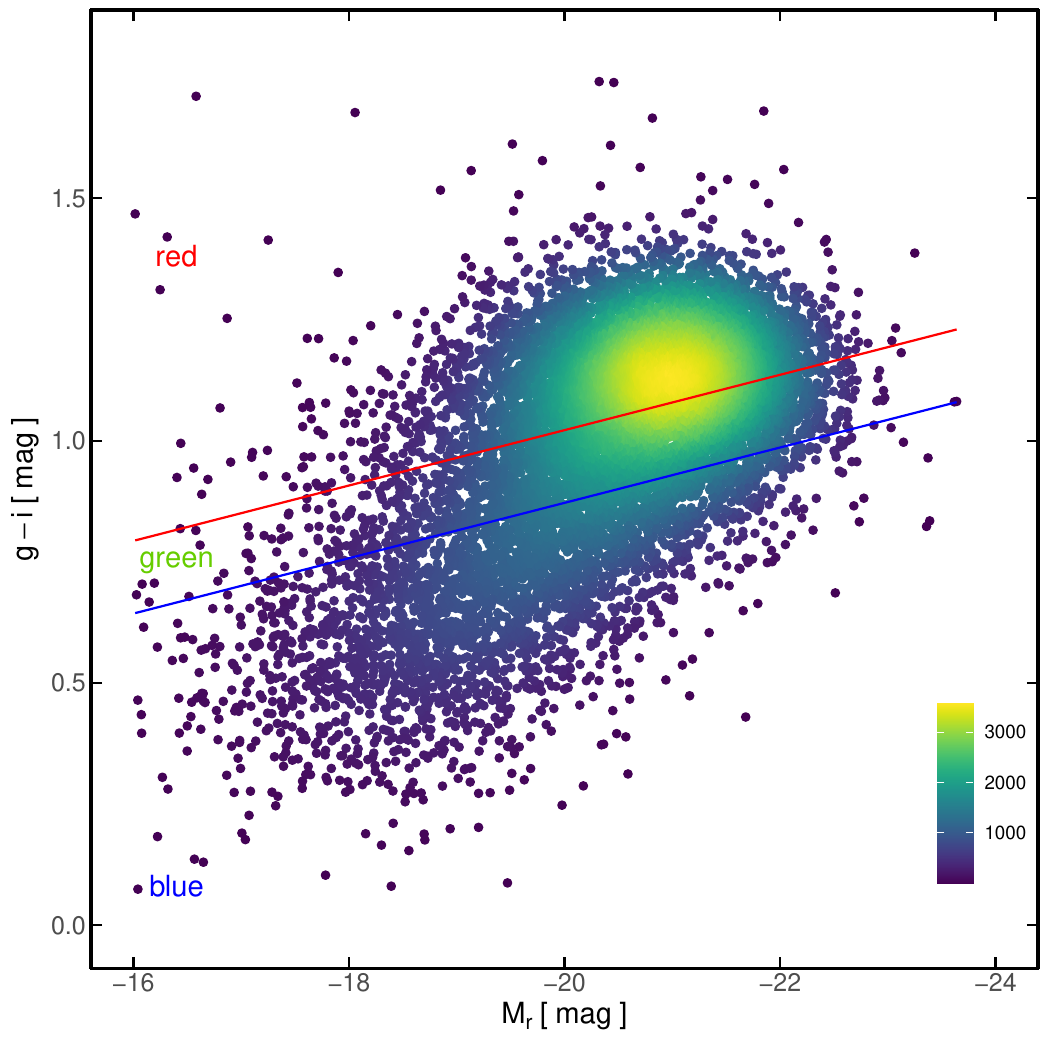}
    \hspace{0.05\textwidth}
    \includegraphics[width=0.36\textwidth]{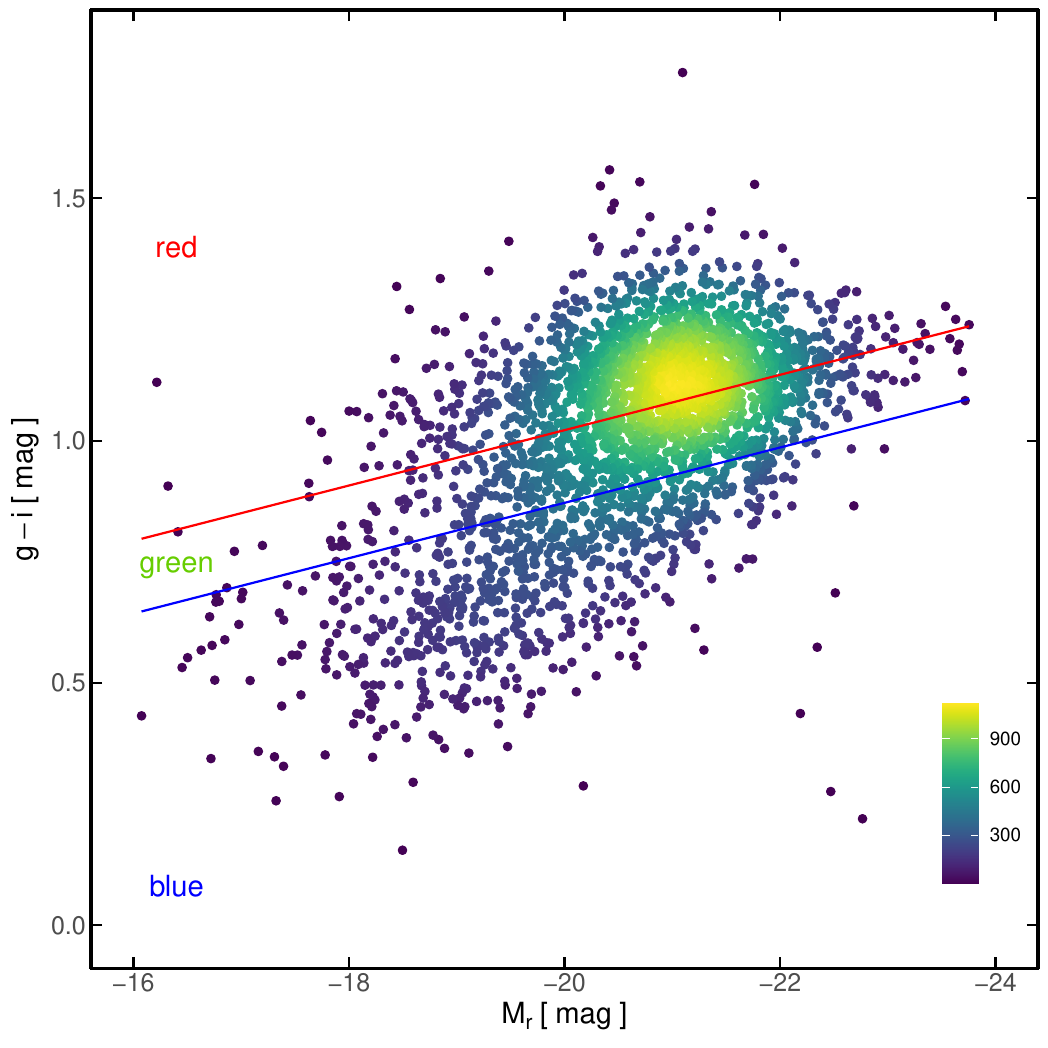}
    \caption{Galaxy colour-magnitude diagram separated into red, green, and blue areas, using only galaxies for which the redshift is known. The colours were corrected for Galactic extinction and $K$ correction as described in Section \ref{sample}. The red line is defined as $g-i=-0.0571(M_r + 24) + 1.2$, while the blue line is offset by 0.15 mag \citep{papastergis_2013}. Left: Sample of detected galaxies with $Confidence>0.7$. Right: Galaxy Zoo Sample used for training.}
    \label{f:3}
\end{figure*}

Looking closer at our sample of galaxies, in Fig. \ref{f:2} we can see that our galaxies are nearby, with most of the redshifts being about $0.02<z<0.10$, which is expected for our sample and is in good agreement with the redshifts of the Galaxy Zoo sample that we used for training. In Fig. \ref{f:3} we show the colour-magnitude diagram (CMD) of our sample compared to the CMD of the training sample obtained from Galaxy Zoo 2 as described in Section \ref{data}. We divide the magnitude into red, green and blue parts, based on criteria from \cite{papastergis_2013}. Most of our galaxies are red, which may seem unexpected, since we are looking for spiral galaxies, but this result is in agreement with \cite{makarov_2022} who made a catalog of $16,551$ edge-on galaxies created using the public DR2 data from the Pan-STARRS survey. They show that edge-on galaxies are redder than general samples of galaxies containing all inclinations, due to internal extinction of edge-on galaxies, which is also why these galaxies are usually excluded when calculating Tully-Fisher relationship \citep[see for example][]{karachentsev_2002}. Moreover, there is excellent agreement between our detected sample and the training sample, showing that our algorithm reproduces the properties of the original sample.

\begin{figure*}[!b]
    \centering
    \includegraphics[width=0.35\textwidth]{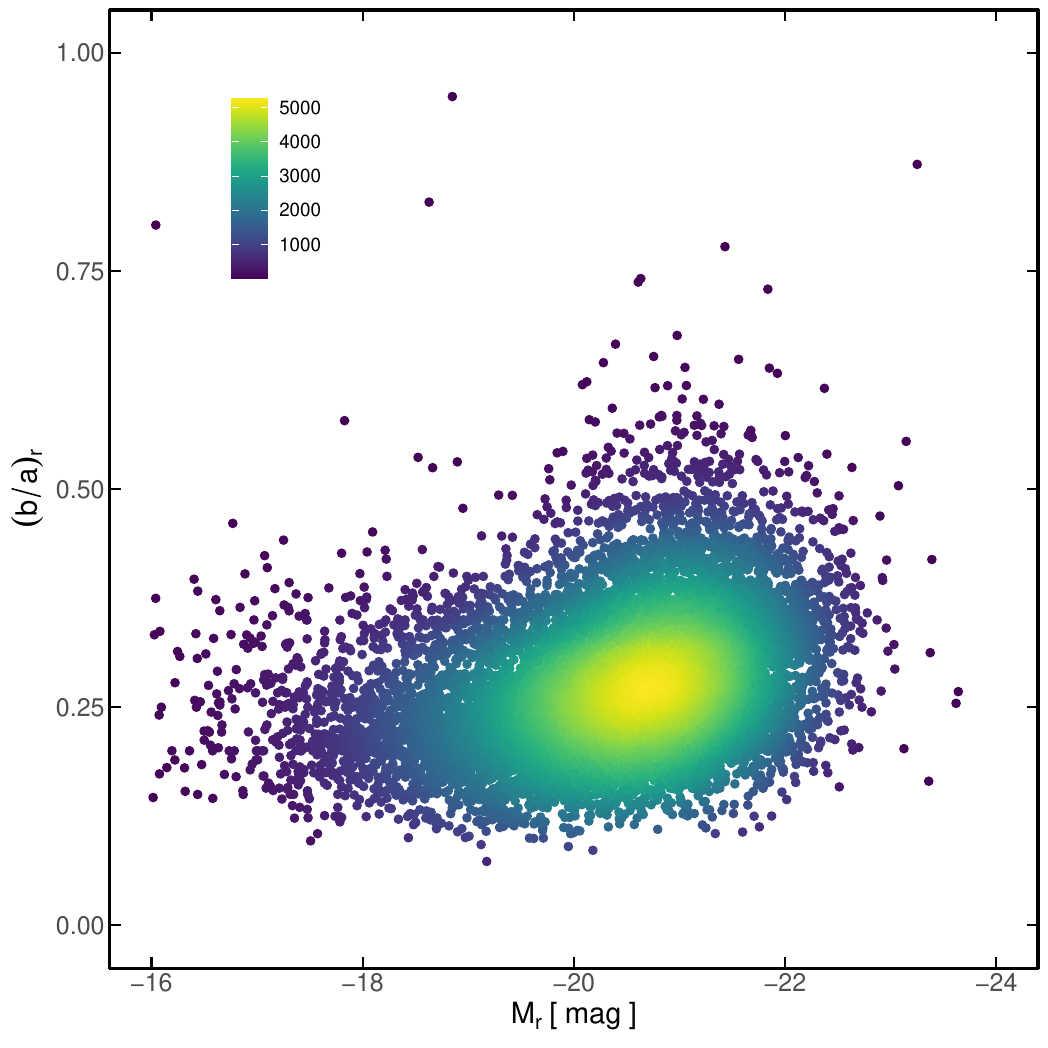}
    \hspace{0.05\textwidth}
    \includegraphics[width=0.35\textwidth]{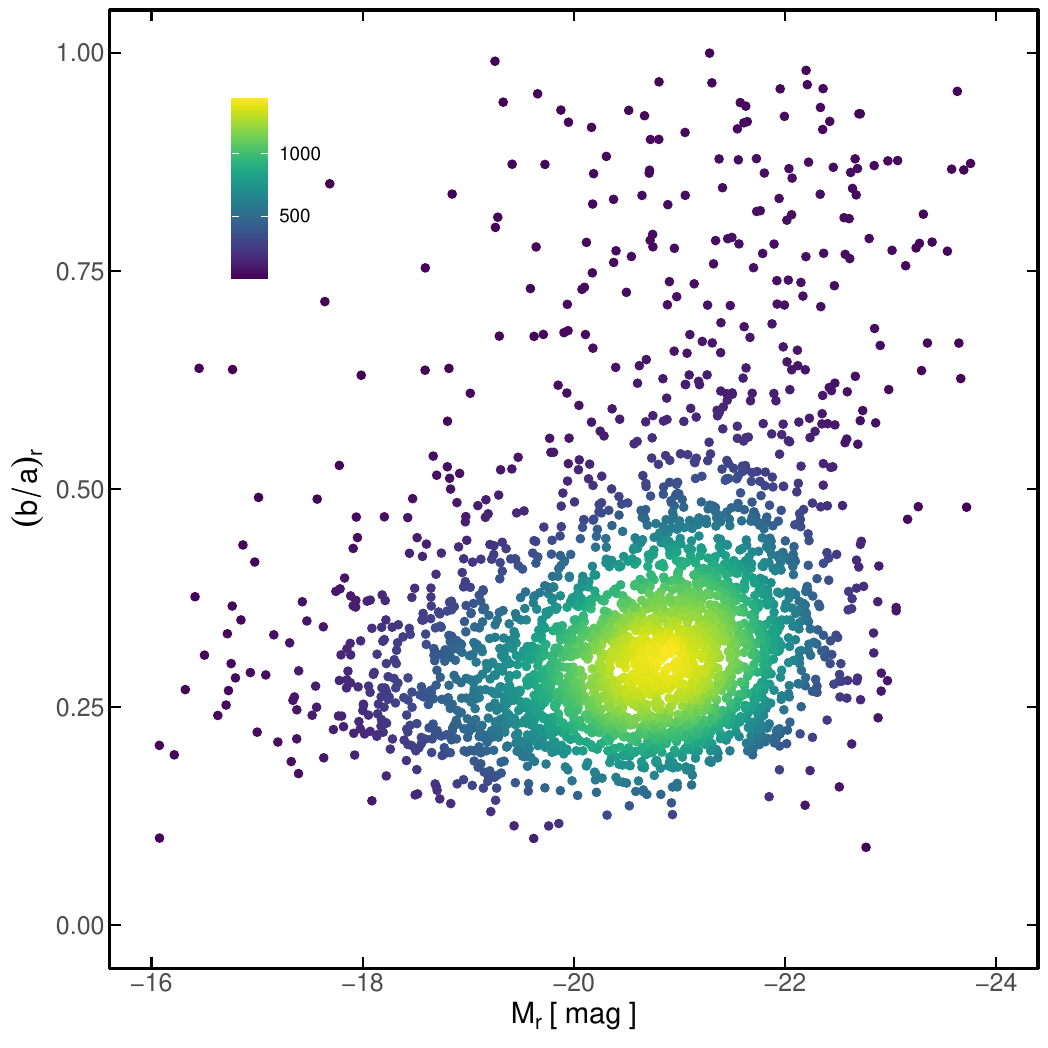}
    \caption{Minor to major axes ratio as a function of absolute magnitude, using only galaxies for which the redshift is known. Left: Sample of detected galaxies with $Confidence>0.7$. Right: Galaxy Zoo Sample used for training.}
    \label{f:4}
\end{figure*}

\begin{figure}
\centering
\includegraphics[width=0.45\textwidth]{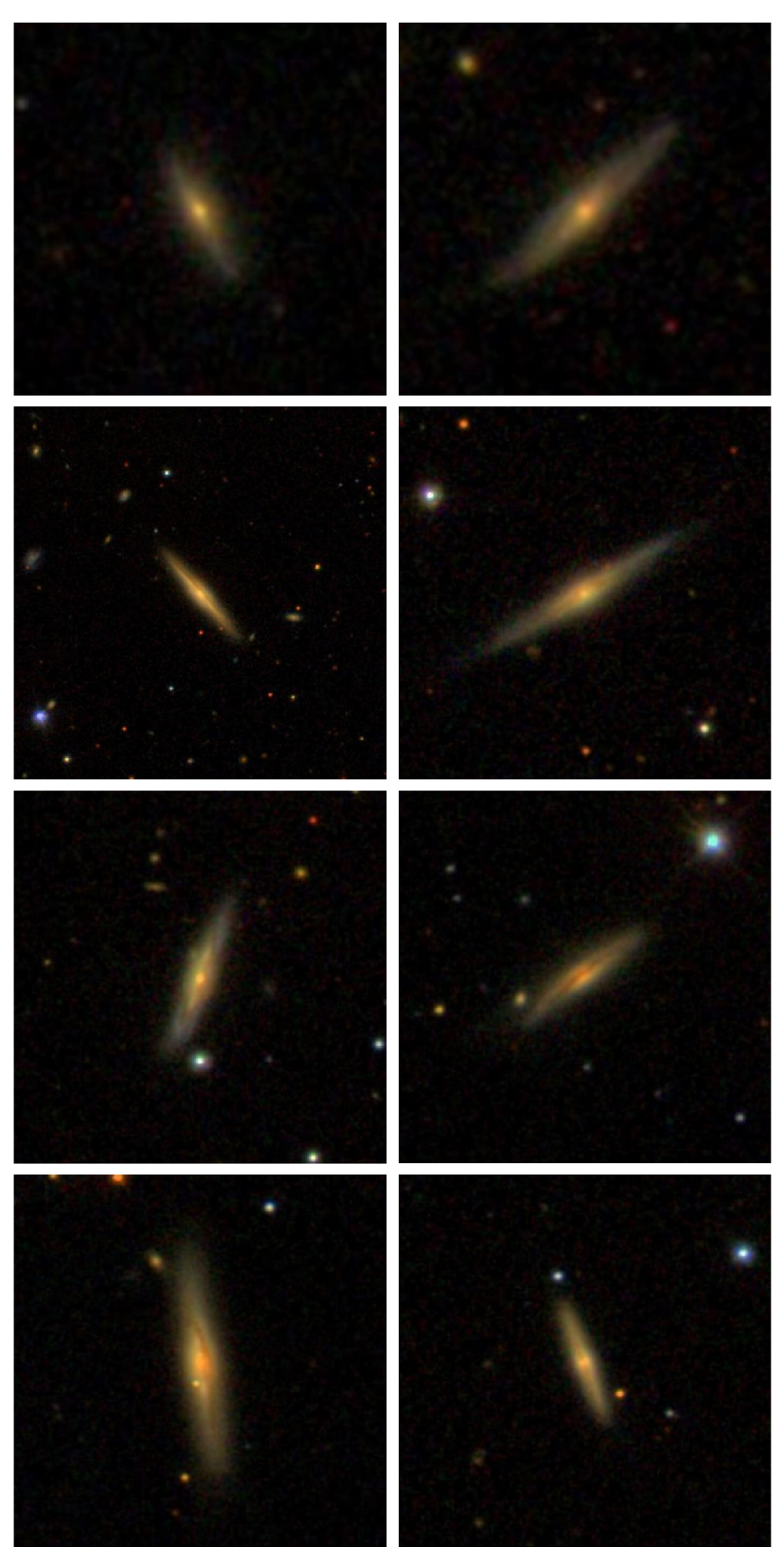}
\caption{Example of several galaxies detected by our algorithm, that have high values of $b/a$ ratio, despite looking like edge-on galaxies. All galaxies shown above have $0.35<(b/a)_r<0.39$.}
\label{f:7}
\end{figure}

The minor to major axes ratio $b/a$, which can be seen in Fig. \ref{f:4} is low for most detected galaxies, which is expected from edge-on galaxies and is in agreement with \cite{makarov_2022}. We see that there are some outliers that seem thicker; however, when we compare our sample with the training sample from Galaxy Zoo, we see that this sample has a similar trend of $b/a$. Therefore, our algorithm is correctly detecting galaxies similar to what it was trained on. Moreover, on average the detected sample has lower values of $b/a$ than the training sample, so our algorithm shows some improvement compared to the training sample, which may suffer from an annotation bias. Another point is that some galaxies that have a relatively high value of $b/a$ in the SDSS catalog appear as thin and edge-on or highly inclined. We illustrate this in Fig. \ref{f:7}, where we show some examples of galaxies with $0.35<(b/a)_r<0.39$, although they appear as thin and edge-on, therefore, the algorithm detected them correctly.

\begin{figure}
    \centering
    \includegraphics[width=0.45\textwidth]{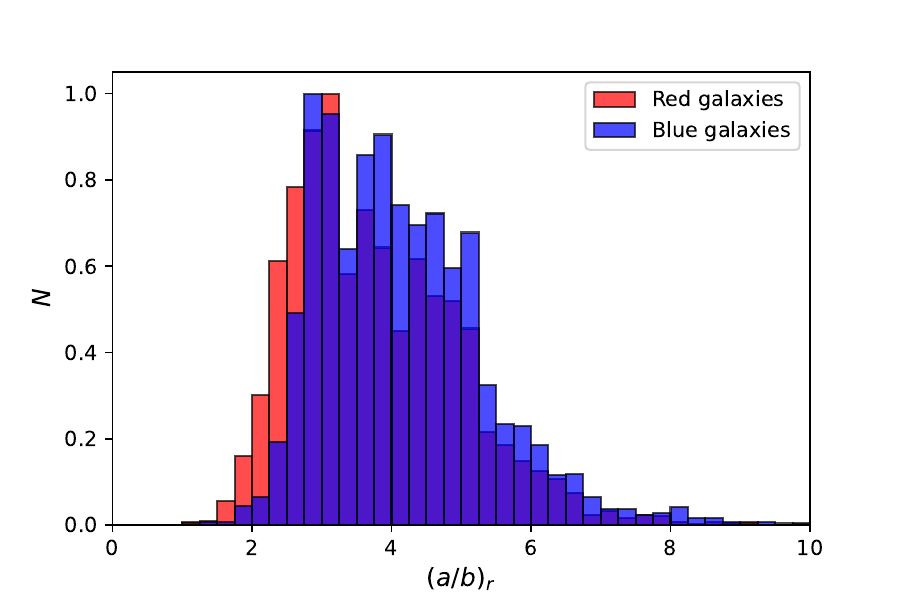}
    \caption{Normalized histogram of the inverse thickness of the detected galaxies for red and blue galaxies. Only galaxies with $Confidence>0.7$ are considered. The galaxies were separated as in Fig. \ref{f:3}.}
    \label{f:6}
\end{figure}

\cite{makarov_2022} also found that on average, red galaxies are thicker than blue ones. In order to corroborate this, we plot a normalized histogram of inverse galaxy thickness in Fig. \ref{f:6} for the red and blue populations separately. We confirm this finding, showing that red galaxies tend to be more thick, while galaxies with lower thickness are more likely to be blue.

We also compared our detected sample to the catalog of \cite{bizyaev_2014} who compiled a catalog of $5,747$ edge-on galaxies from the SDSS DR7 supplemented by data from other surveys (Revised Flat Galaxy Catalog \citep{karachentsev_1999}, RC3 \citep{de_vaucouleurs_1991}, EFIGI \citep{baillard_2011}, and Galaxy Zoo \citep{lintott_2011}), selected by visual classification. A simple cross-match of our samples reveals that we detected approximately $3,500$ identical galaxies. Considering that we use two different SDSS data releases and that \cite{bizyaev_2014} also added galaxies from a few different catalogs, we find this agreement reasonable. Moreover, as we showed, we found a good agreement of the properties of our sample with previous studies.

\section{Conclusions}\label{concl}
We present a novel deep learning algorithm for the detection and extraction of edge-on galaxies in astronomical images. Based on Galaxy Zoo classification, we obtained images of nearly $16,000$ edge-on, spiral galaxies from the SDSS catalog that we trained the algorithm on. Our results show that the algorithm can reliably detect edge-on galaxies, with Precision=0.80 and Recall=0.94. We also successfully applied a segmentation algorithm to extract the galaxies from the image, so they can be used for future studies. We retrieved the parameters of our sample of galaxies from the SDSS and provide them together with confidence values. From a basic analysis of the properties of our sample, we find that the majority of galaxies are nearby ($0.02<z<0.10$), red and have low values of $b/a$, which is in agreement with other works, as well as the training data. Our analysis of the SDSS data showed that we can reliably detect the needed galaxies, however, the true potential of the algorithm lies in its possible application on data from future surveys, such as Euclid \citep{euclid_2024}, Roman Space Telescope \citep{bartusek_2022} or Large Synoptic Survey Telescope \citep[LSST,][]{ivezic_2019} where we expect a significant increase in data volume, which necessitates processing through machine learning. In a future work, we will develop this algorithm to automatically detect warped galaxies.

\begin{acknowledgments}
We thank Daisuke Kawata, Ralph Sch\"{o}nrich and Mat Page for useful advice and comments. We thank Ani Thakar for facilitating the SDSS data download. Z.C. was supported by Royal Society grant No. UF160158/ RF\textbackslash ERE\textbackslash210415. V.K. and P.B. were supported by VEGA-the Slovak Grant Agency for Science, grant No. 1/0685/21. R.N. was funded by the EU NextGenerationEU through the Recovery and Resilience Plan for Slovakia under the project No. 09I03-03-V04-00137. This research made use of the ``K-corrections calculator'' service available at http://kcor.sai.msu.ru/. Funding for the Sloan Digital Sky Survey (SDSS) and SDSS-II has been provided by the Alfred P. Sloan Foundation, the Participating Institutions, the National Science Foundation, the U.S. Department of Energy, the National Aeronautics and Space Administration, the Japanese Monbukagakusho, and the Max Planck Society, and the Higher Education Funding Council for England. The SDSS Web site is http://www.sdss.org/.
The SDSS is managed by the Astrophysical Research Consortium (ARC) for the Participating Institutions. The Participating Institutions are the American Museum of Natural History, Astrophysical Institute Potsdam, University of Basel, University of Cambridge, Case Western Reserve University, The University of Chicago, Drexel University, Fermilab, the Institute for Advanced Study, the Japan Participation Group, The Johns Hopkins University, the Joint Institute for Nuclear Astrophysics, the Kavli Institute for Particle Astrophysics and Cosmology, the Korean Scientist Group, the Chinese Academy of Sciences (LAMOST), Los Alamos National Laboratory, the Max-Planck-Institute for Astronomy (MPIA), the Max-Planck-Institute for Astrophysics (MPA), New Mexico State University, Ohio State University, University of Pittsburgh, University of Portsmouth, Princeton University, the United States Naval Observatory, and the University of Washington.
This publication uses data generated via the Zooniverse.org platform, development of which is funded by generous support, including a Global Impact Award from Google, and by a grant from the Alfred P. Sloan Foundation.
\end{acknowledgments}

%

\vspace{5mm}
\facilities{Sloan}


\software{APLpy \citep{aplpy2012,aplpy2019}, astropy \citep{astropy_2013, astropy_2018, astropy_2022}, astroquery \citep{astroquery_2019}, dustmaps \citep{green_2018}, ggpointdensity (\url{https://github.com/LKremer/ggpointdensity}), K-corrections calculator \citep{chilingarian_2010, chilingarian_2012}, matplotlib \citep{Hunter_2007}, pandas \citep{mckinney_2010, pandas_2024}, SciServer \citep{sciserver_2020}}



\bibliography{Refer}{}
\bibliographystyle{aasjournal}



\end{document}